\def\ra{\rangle}

\def\be{\begin{equation}}
\def\ee{\end{equation}}
\def\ba{\begin{array}}
\def\ea{\end{array}}

\documentclass[11pt]{article}
\usepackage{bbm}
\usepackage{amsfonts}
\usepackage{mathrsfs}
\usepackage{pifont}
\usepackage{amsmath}
\usepackage{amssymb}
\usepackage{caption2}
\usepackage{graphicx}
\begin{document}

\title{A Note on Mutually Unbiased Unextendible Maximally Entangled Bases in $\mathbb{C}^{2}\bigotimes \mathbb{C}^{3}$}

\author{ Halqem Nizamidin$^{1}$, Teng Ma$^{1}$  and Shao-Ming Fei$^{1,2}$ \\
{\footnotesize  {$^1$School of Mathematical Sciences, Capital Normal
University, Beijing 100048, China}}\\
{\footnotesize{$^2$Max-Planck-Institute for Mathematics in the
Sciences, 04103 Leipzig, Germany}}}

\date{}
\maketitle

\centerline{\bf Abstract}
\medskip

We systematically study the construction of
mutually unbiased bases in $\mathbb{C}^{2}\bigotimes\mathbb{C}^{3}$, such that
all the bases are unextendible maximally entangled ones. Necessary conditions of constructing a
pair of mutually unbiased unextendible maximally entangled bases in $\mathbb{C}^{2}\bigotimes\mathbb{C}^{3}$ are derived.
Explicit examples are presented.

\bigskip
\medskip

Mutually unbiased bases (MUBs) play important roles in many quantum information processing
such as quantum state tomography \cite{s5,qst1,qst2}, cryptographic
protocols \cite{cp1,cp2}, and the mean king¡¯s problem \cite{mk}.
They are also useful in the construction of generalized Bell states.
Let $\mathcal{B}_{1}=\{|\phi_{i}\rangle\}$ and $\mathcal{B}_{2}=\{|\psi_{i}\rangle\}$, $i=1,2,\cdots,d$,
be two orthonormal bases of a $d$-dimensional complex vector space $\mathbb{C}^{d}$,
$\langle\phi_j|\phi_{i}\rangle=\delta_{ij}$, $\langle\psi_j|\psi_{i}\rangle=\delta_{ij}$.
$\mathcal{B}_{1}$ and $\mathcal{B}_{2}$ are said to be mutually unbiased if and only if
\begin{equation}\label{mu}
|\langle\phi_{i}|\psi_{j}\rangle| =\frac{1}{\sqrt{d}}~~~\forall~ i,j=1,2,\cdots,d.
\end{equation}
Physically if a system is prepared in an eigenstate of basis $\mathcal{B}_{1}$
and is measured in basis $\mathcal{B}_{2}$, then all the measurement outcomes have the same probability.

A set of orthonormal bases $\{\mathcal{B}_{1}, \mathcal{B}_{2},...,\mathcal{B}_{m}\}$ in $\mathbb{C}^{d}$
is called a set of MUBs if every pair of bases in the set is mutually unbiased.
For given dimensional $d$, the maximum number of MUBs is no more than $d+1$.
It has been shown that there are $d+1$ MUBs when $d$ is a prime power \cite{s5,I.
D. Ivanovic, T. Durt}.  However, for general $d$, e.g. $d=6$, it is a formidable problem to determine
the maximal numbers of MUBs \cite{s6,s7,s8,s9,P. Butterley,S. Brierley3,P. Jaming,T. Paterek,I. Bengtsson}.

When the vector space is a bipartite system $\mathbb{C}^{d}\bigotimes\mathbb{C}^{d'}$ of composite dimension $dd'$,
there are different kinds of bases in $\mathbb{C}^{d}\bigotimes\mathbb{C}^{d'}$ according to the entanglement
of the basis vectors.
The unextendible product basis (UPB) is a set of incomplete orthonormal
product basis whose complementary space has no product states \cite{s2}.
It is shown that the mixed state on the subspace complementary
to a UPB is a bound entangled state. Moreover,
the states comprising a UPB are not distinguishable by local measurements and classical communication.

The unextendible maximally entangled basis (UMEB) is a set of orthonormal maximally entangled states in
$\mathbb{C}^{d}\bigotimes\mathbb{C}^{d}$ consisting of less than $d^2$ vectors which have no additional maximally
entangled vectors that are orthogonal to all of them \cite {s3}.
Recently, the UMEB in arbitrary bipartite spaces $\mathbb{C}^{d}\bigotimes\mathbb{C}^{d'}$
has be investigated in \cite{Bin Chen}. A systematic way in constructing
$d^{2}$-member UMEBs in $\mathbb{C}^{d}\bigotimes\mathbb{C}^{d'}$ ($\frac{d'}{2}<d<d'$)
is presented. It is shown that the subspace complementary to the
$d^{2}$-member UMEB contains no states of Schmidt rank higher than
$d-1$. From the approach of constructing UMEBs in $\mathbb{C}^{d}\bigotimes\mathbb{C}^{d'}$,
two mutually unbiased UMEBs in $\mathbb{C}^{2}\bigotimes\mathbb{C}^{3}$ are constructed in \cite{Bin Chen}.

In this note, we systematically study the UMEBs in $\mathbb{C}^{2}\bigotimes\mathbb{C}^{3}$
and present a generic way in constructing a pair of
UMEBs in $\mathbb{C}^{2}\bigotimes\mathbb{C}^{3}$ such that they are mutually unbiased. The special
example given in \cite{Bin Chen} can be easily obtained from our approach.

A set of states $\{|\phi_{i}\rangle\}$ in  $\mathbb{C}^{d}\bigotimes\mathbb{C}^{d'}$, $i=1,2,\cdots n$, $n<dd'$, is called an $n$-member UMEB if and
only if

(i) all the states $|\phi_{i}\rangle$ are maximally entangled;

(ii) $\langle\phi_{i}|\phi_{j}\rangle=\delta_{i,j}$;

(iii) if $\langle\phi_{i}|\psi\rangle=0$, $\forall~i=1,2,\cdots,n$, then $|\psi\rangle$ cannot be maximally entangled.

Here a state $|\psi\rangle$ is said to be a
$\mathbb{C}^{d}\bigotimes\mathbb{C}^{d'}$ maximally entangled state
if and only if for an arbitrary given orthonormal complete basis
$\{|i_{A}\rangle\}$ of the subsystem $A$, there exist an orthonormal
basis $\{|i_{B}\rangle\}$ of the subsystem $B$ such that  $|\psi\rangle$
can be written as $|\psi\rangle=\frac{1}{\sqrt{d}}\Sigma_{i=0}^{d-1}|i_{A}\rangle\otimes|i_{B}\rangle$ \cite{Z.-G. Li}.

Let $\{|0\rangle, |1\rangle\}$ and $\{|0'\rangle, |1'\rangle\,
|2'\rangle\}$ be the computational bases in $\mathbb{C}^{2}$ and
$\mathbb{C}^{3}$ respectively. To construct a pair of MUBs
which are both UMEBs in $\mathbb{C}^{2}\bigotimes\mathbb{C}^{3}$,
we start with the first UMEB in $\mathbb{C}^{2}\bigotimes\mathbb{C}^{3}$
given by
\begin{eqnarray}\label{frist UMEB bases}
&& |\phi_{i}\rangle=\frac{1}{\sqrt{2}}(\sigma_{i}\otimes I_{3})(|00'\rangle+|11'\rangle),\nonumber\\
&& |\phi_{4}\rangle=|0\rangle\otimes|2'\rangle,\nonumber\\
&& |\phi_{5}\rangle=|1\rangle\otimes|2'\rangle,
\end{eqnarray}
where $\sigma_{0}$ denotes the $2\times 2$ identity matrix, $\sigma_{i}$, $i=1,2,3$, are the Pauli matrices,
$I_{3}$ stands for the $3\times 3$ identity matrix, $|\alpha\beta\rangle\equiv |\alpha\rangle\otimes |\beta\rangle$.

If we choose $\{|a\rangle, |b\rangle\}$ and
$\{|x'\rangle, |y'\rangle, |z'\rangle\}$ to be another two bases of $\mathbb{C}^{2}$  and
$\mathbb{C}^{3}$ respectively, then we have the second UMEB in $\mathbb{C}^{2}\bigotimes \mathbb{C}^{3}$,
\begin{eqnarray}\label{second UMEB bases}
&&|\psi_{i}\rangle=\frac{1}{\sqrt{2}}(\sigma_{i}\otimes I_{3})(|0x'\rangle+|1y'\rangle)\nonumber,\\
&&|\psi_{4}\rangle=|a\rangle\otimes|z'\rangle\nonumber,\\
&&|\psi_{5}\rangle=|b\rangle\otimes|z'\rangle.
\end{eqnarray}

The bases $\{|\phi_{i}\rangle\}$ and $\{|\psi_{i}\rangle\}$ are mutually unbiased
if and only if they satisfy the relations (\ref{mu}),
\begin{equation}\label{mu23}
|\langle\phi_{i}|\psi_{j}\rangle|=\frac{1}{\sqrt{6}}, ~~\forall~i,j=0,1,\cdots,5.
\end{equation}

Let $S$ and $W$ be the unitary matrixes that transforms the bases $\{|0\rangle,|1\rangle\}$ and
$\{|0'\rangle,|1'\rangle\,|2'\rangle\}$ to $\{|a\rangle,|b\rangle\}$ and $\{|x'\rangle,|y'\rangle,|z'\rangle\}$
respectively,
\begin{eqnarray}\label{5}
&&S(|0\rangle,|1\rangle)=(|a\rangle,|b\rangle),\nonumber\\
&&W(|0'\rangle,|1'\rangle,|2'\rangle)=(|x'\rangle,|y'\rangle,|z'\rangle).
\end{eqnarray}
Correspondingly we have the relations between $|\phi_{i}\rangle$ and $|\psi_{j}\rangle$,
\begin{eqnarray}\label{psiphi}
&&|\psi_{j}\rangle=(I_{2}\otimes W)|\phi_{j}\rangle,~~~\forall ~j=0,1,2,3,\nonumber\\
&&|\psi_{j}\rangle=(S\otimes W)|\phi_{j}\rangle,~~~\forall~j=4,5.
\end{eqnarray}
From (\ref{mu23}) one gets,
\begin{equation}\label{rel}
\begin{array} {l}
|\langle\phi_{i}|I_{2}\otimes
W|\phi_{j}\rangle| =\frac{1}{\sqrt{6}},  ~\forall ~i=0,1,...,5,~ j=0,1,2,3, \\
|\langle\phi_{i}|S\otimes W|\phi_{j}\rangle|
=\frac{1}{\sqrt{6}},~\forall ~i=0,1,...,5,~j=4,5.\\
\end{array}
\end{equation}

As $\{|\phi_{i}\rangle\}$ forms a base in $\mathbb{C}^{2}\bigotimes \mathbb{C}^3$, the relations
in (\ref{rel}) imply that the absolute values of the entries of the matrices $I\otimes  W$
and $S\otimes W$ under the base $\{|\phi_{i}\rangle\}$ have the following forms:
\begin{equation}\label{x1}
\left(\begin{array}{cccccc}
\frac{1}{\sqrt{6}}&\frac{1}{\sqrt{6}}&\frac{1}{\sqrt{6}}&\frac{1}{\sqrt{6}}&X&X\\
\frac{1}{\sqrt{6}}&\frac{1}{\sqrt{6}}&\frac{1}{\sqrt{6}}&\frac{1}{\sqrt{6}}&X&X\\
\frac{1}{\sqrt{6}}&\frac{1}{\sqrt{6}}&\frac{1}{\sqrt{6}}&\frac{1}{\sqrt{6}}&X&X\\
\frac{1}{\sqrt{6}}&\frac{1}{\sqrt{6}}&\frac{1}{\sqrt{6}}&\frac{1}{\sqrt{6}}&X&X\\
\frac{1}{\sqrt{6}}&\frac{1}{\sqrt{6}}&\frac{1}{\sqrt{6}}&\frac{1}{\sqrt{6}}&X&X\\
\frac{1}{\sqrt{6}}&\frac{1}{\sqrt{6}}&\frac{1}{\sqrt{6}}&\frac{1}{\sqrt{6}}&X&X
\end{array}
\right),
\end{equation}
\begin{equation}\label{x2}
 \left(\begin{array}{cccccc}
X&X&X&X&\frac{1}{\sqrt{6}}&\frac{1}{\sqrt{6}}\\
X&X&X&X&\frac{1}{\sqrt{6}}&\frac{1}{\sqrt{6}}\\
X&X&X&X&\frac{1}{\sqrt{6}}&\frac{1}{\sqrt{6}}\\
X&X&X&X&\frac{1}{\sqrt{6}}&\frac{1}{\sqrt{6}}\\
X&X&X&X&\frac{1}{\sqrt{6}}&\frac{1}{\sqrt{6}}\\
X&X&X&X&\frac{1}{\sqrt{6}}&\frac{1}{\sqrt{6}}
\end{array}
\right),
\end{equation}
where X denotes any numbers.

Let
\begin{equation}
S= \left(
       \begin{array}{cc}
         s_{11} & s_{12} \\
         s_{21}& s_{22} \\
       \end{array}
     \right), ~~
W=\left(
        \begin{array}{ccc}
          w_{11} & w_{12} & w_{13} \\
          w_{21}& w_{22} & w_{23}\\
          w_{31} & w_{32}& w_{33} \\
        \end{array}
      \right)
\end{equation}
be the matrices of $S$ and $W$ in the computational product basis
$\{|0\rangle,|1\rangle\}\otimes\{|0'\rangle,|1'\rangle,|2'\rangle\}$.
Let $F$ be the unitary matrix  that transforms the computational product basis to
the basis $\{|\phi_{i}\rangle\}$, i.e.,$
F(|00'\ra, |01'\ra, |02'\ra, |10'\ra, |11'\ra, |12'\ra)=(|\phi_{0}\rangle,...,|\phi_{5}\rangle)$.
Form (2), one can easily get
\begin{equation}
 F=\left(\begin{array}{cccccc}
\frac{1}{\sqrt{2}}&0&0&\frac{1}{\sqrt{2}}&0&0\\
0&\frac{1}{\sqrt{2}}&-\frac{1}{\sqrt{2}}&0&0&0\\
0&0&0&0&1&0\\
0&\frac{1}{\sqrt{2}}&\frac{1}{\sqrt{2}}&0&0&0\\
\frac{1}{\sqrt{2}}&0&0&-\frac{1}{\sqrt{2}}&0&0\\
0&0&0&0&0&1\\
\end{array}
\right).
\end{equation}

Therefore the matrices of $I_2 \otimes W$ and
$S\otimes W$ under the basis $\{|\phi_{i}\rangle\}$ are given by
\begin{equation}\label{xx1}
F^{\dagger}(I_{2}\otimes W)F,
\end{equation}
and
\begin{equation}\label{xx2}
F^{\dagger}(S\otimes W)F,
\end{equation}
respectively.

Comparing (\ref{xx1}) and (\ref{xx2}) with (\ref{x1}) and (\ref{x2}), by straightforward
calculations, we have

(i) The absolute values of the entries of $w$ are $1/\sqrt{3}$. Moreover, in the complex plane,
$w_{11}\perp w_{22}$ and  $w_{21}\perp w_{12}$.

(ii) The absolute values of the entries of $S$ is $1/\sqrt{2}$. In
the complex plane, $w_{13}s_{11}\perp w_{23}s_{21}$,
$w_{23}s_{11}\perp w_{13}s_{21}$, $w_{13}s_{12}\perp w_{23}s_{22}$ and
$w_{23}s_{12}\perp w_{13}s_{22}$.

From the condition (i), for simplification, we can set
\begin{equation}\label{Wc}
W=1/\sqrt{3}\begin{pmatrix}
e^{i\theta_{1}}& e^{i(\theta_{2}+\frac{\pi}{2})}&e^{i\theta_{4}}\\
e^{i\theta_{2}}& e^{i(\theta_{1}+\frac{\pi}{2})}&e^{i\theta_{5}}\\
e^{i\theta_{3}}& e^{i(\theta_{3}-\frac{\pi}{2})}&e^{i\theta_{6}}
\end{pmatrix},
\end{equation}
where, due the properties of unitary matrix, $\theta_i$ satisfy the following conditions,
\begin{equation}\label{condiw}
\begin{array}{l}
|\theta_{1}-\theta_{2}|=\frac{\pi}{3},~~~|\theta_{4}-\theta_{5}|=\pi,\\[1mm]
e^{i(\theta_{1}-\theta_4)}e^{-i\pi/3}+e^{i(\theta_{3}-\theta_{6})}=0.
\end{array}
\end{equation}

From equation (\ref{Wc}), (\ref{condiw}) and condition (ii), we find $s_{11}$ and $s_{21}$ are orthogonal, $s_{12}$ and $s_{22}$ are orthogonal. Then we can simply  set
\begin{equation}\label{condis}
S=\frac{1}{\sqrt{2}}\begin{pmatrix}
e^{i\theta_{1}'}&e^{i\theta_{2}'}\\
\pm e^{i(\theta_{1}'+\frac{\pi}{2})}&\mp e^{i(\theta_{2}'+\frac{\pi}{2})}\\
\end{pmatrix}.
\end{equation}
where  $\theta_{1}',\theta_{2}'$ can be any real numbers.

Therefore, for any $\theta_{i}$s and $\theta_{i}'$s satisfying (\ref{condiw}) and (\ref{condis})
respectively, one has a $W$ and a $S$. Then from (\ref{psiphi}) one gets the UMEB
$\{|\psi_{i}\rangle\}$ that is mutually unbiased with the UMEB $\{|\phi_{i}\rangle\}$.

We next give some concrete examples of mutually unbiased UMEBs in $\mathbb{C}^{2}\bigotimes \mathbb{C}^{3}$.

The UMEB $\{|\phi_{i}\rangle\}$ presented in \cite{Bin Chen} is of the form,
\begin{eqnarray}\label{UMEB1'}
&& |\phi_{0}\rangle=\frac{1}{\sqrt{2}}(|00'\rangle+|11'\rangle)\nonumber,\\
&& |\phi_{i}\rangle=\frac{1}{\sqrt{2}}(\sigma_{i}\otimes I_{3})(|00'\rangle+|11'\rangle),~~~i=1,2,3,\nonumber\\
&& |\phi_{4}\rangle=|c\ra\otimes|2'\rangle,\nonumber\\
&& |\phi_{5}\rangle=|d\ra\otimes|2'\rangle,
\end{eqnarray}
where $|c\ra=\frac{1}{2}|0\ra+\frac{\sqrt{3}}{2}|1\ra,|d\ra=\frac{\sqrt{3}}{2}|0\ra-\frac{1}{2}|1\ra.$
This example corresponds to a different transformation matrix $F$,
\begin{equation}\nonumber
 F=\left(\begin{array}{cccccc}
\frac{1}{\sqrt{2}}&0&0&\frac{1}{\sqrt{2}}&0&0\\
0&\frac{1}{\sqrt{2}}&-\frac{1}{\sqrt{2}}&0&0&0\\
0&0&0&0&\frac{1}{2}&\frac{\sqrt{3}}{2}\\
0&\frac{1}{\sqrt{2}}&\frac{1}{\sqrt{2}}&0&0&0\\
\frac{1}{\sqrt{2}}&0&0&-\frac{1}{\sqrt{2}}&0&0\\
0&0&0&0&\frac{\sqrt3}{2}&-\frac{1}{2}\\
\end{array}
\right).
\end{equation}
From our approach, $w_{31}\perp w_{32}$ should be added to the condition (i).
With respect to the condition (ii), the orthogonal relation becomes
$(s_{11}+\sqrt{3}s_{12})\perp (s_{21}+\sqrt{3}s_{22})$ and $(\sqrt{3}s_{11}-s_{12})\perp (\sqrt{3}s_{21}-s_{22})$.
However, since we have already set $w_{31}\perp w_{32}$ in (\ref{Wc}), (\ref{condiw}) can be also used for this example.

We choose $\{\theta_{i}\}$ to be
\begin{equation}\label{exam}
\{\theta_{1}=0,\,\theta_{2}=\frac{\pi}{3},\, \theta_{3}=0,\theta_{4}=\pi ,\theta_{5}=0,\,\theta_{6}=\frac{\pi}{3}\},
\end{equation}
which satisfy the condition (\ref{condiw}). From (\ref{Wc}) we have
\begin{equation}
W=1/\sqrt{3}
\begin{pmatrix}
1&\frac{-\sqrt{3}+i}{2}&-1\\
\frac{1+\sqrt{3}i}{2}&i&1\\
1&-i&\frac{1+\sqrt{3}i}{2}\\
\end{pmatrix}.
\end{equation}

The unitary matrix $W$ transforms the basis $\{|0'\rangle,|1'\rangle\,|2'\rangle\}$ to
basis $\{|x'\rangle,|y'\rangle,|z'\rangle\}$. From (\ref{5}) we have
\begin{eqnarray}\label{x,y,z}
&&|x'\rangle=\frac{1}{\sqrt{3}}(|0'\rangle+\frac{1+\sqrt{3}i}{2}|1'\rangle+|2'\rangle)\nonumber,\\
&&|y'\rangle=\frac{1}{\sqrt{3}}(\frac{-\sqrt{3}+i}{2}|0'\rangle+i|1'\rangle-i|2'\rangle)\nonumber,\\
&&|z'\rangle=\frac{1}{\sqrt{3}}(-|0'\rangle+|1'\rangle+\frac{1+\sqrt{3}i}{2}|2'\rangle).
\end{eqnarray}

We have the unitary operator $S$,
\begin{equation}\label{condis'}
S=\frac{1}{\sqrt{2}}\begin{pmatrix}
1&i\\
\frac{\sqrt{3}+i}{2}&\frac{1-\sqrt{3}\,i}{2}
\end{pmatrix}.
\end{equation}
The corresponding operator $S$, $S(|c\rangle,|d\rangle)=(|a\rangle,|b\rangle)$,
give rise to
\begin{eqnarray}\label{a,b}
&&|a\rangle=\frac{1}{\sqrt{2}}(\frac{1+\sqrt{3}i}{2}|0\rangle+\frac{\sqrt{3}-i}{2}|1\rangle),\nonumber\\
&&|b\rangle=\frac{1}{\sqrt{2}}(\frac{\sqrt{3}-i}{2}|0\rangle+\frac{1+\sqrt{3}i}{2}|1\rangle).
\end{eqnarray}

Therefore, the second UMEB that is mutually unbiased to (\ref{UMEB1'}) is given by
\begin{eqnarray}\label{UMEB2'}
&& |\psi_{j}\rangle=\frac{1}{\sqrt{2}}(\sigma_{i}\otimes I_{3})(|0x'\rangle+|1y'\rangle),~~~j=1,2,3,\nonumber\\
&&|\psi_{4}\rangle=\frac{1}{\sqrt{2}}(\frac{1+\sqrt{3}i}{2}|0\rangle+\frac{\sqrt{3}-i}{2}|1\rangle)\otimes|z'\rangle,\nonumber\\
&&|\psi_{5}\rangle=\frac{1}{\sqrt{2}}(\frac{\sqrt{3}-i}{2}|0\rangle+\frac{1+\sqrt{3}i}{2}|1\rangle)\otimes|z'\rangle.
\end{eqnarray}
(\ref{UMEB1'}) and (\ref{UMEB2'}) are exactly the ones presented in \cite{Bin Chen}.

Now we give a new example by choosing other values of $\{\theta_{i}\}$ and $\{\theta_i'\}$.
Let the first UMEB in $\mathbb{C}^{2}\bigotimes\mathbb{C}^{3}$ be the one
given in (\ref{frist UMEB bases}). Taking into the condition (\ref{condiw}),
we set
\begin{equation}\label{exam}
\theta_{1}=\pi,\,\theta_{2}=\frac{2\pi}{3},\,\theta_{3}=\theta_{4}=0,\,\theta_{5}=\pi,\,\theta_{6}=\frac{\pi}{3}.
\end{equation}
From (\ref{Wc}), we get
\begin{equation}
W=1/\sqrt{3}\begin{pmatrix}
-1&\frac{-\sqrt{3}-i}{2}&1\\
\frac{-1+\sqrt{3}i}{2}& -i&-1\\
1&-i&\frac{1+\sqrt{3}i}{2}\\
\end{pmatrix},
\end{equation}
and
\begin{eqnarray}\label{x,y,z}
&&|x'\rangle=\frac{1}{\sqrt{3}}(-|0'\rangle+\frac{-1+\sqrt{3}i}{2}|1'\rangle+|2'\rangle)\nonumber,\\
&&|y'\rangle=\frac{1}{\sqrt{3}}(\frac{-\sqrt{3}-i}{2}|0'\rangle-i|1'\rangle-i|2'\rangle)\nonumber,\\
&&|z'\rangle=\frac{1}{\sqrt{3}}(|0'\rangle-|1'\rangle+\frac{1+\sqrt{3}i}{2}|2'\rangle).
\end{eqnarray}

Taking $\theta_{1}'=0$ and $\theta_{2}'=\frac{\pi}{2}$,  we have
\begin{equation}\label{condis''}
S=\frac{1}{\sqrt{2}}\begin{pmatrix}
1&i\\
i&1\\
\end{pmatrix},
\end{equation}
and
\begin{equation}\label{a,b}
|a\rangle=\frac{1}{\sqrt{2}}(|0\rangle+i|1\rangle),~~~~
|b\rangle=\frac{1}{\sqrt{2}}(i|0\rangle+|1\rangle).
\end{equation}

From (\ref{second UMEB bases}) we obtain the second UMEB that is
mutually unbiased to the UMEB given by Eq. (2),
\begin{eqnarray}\label{UMEB2'p}
&& |\psi_{j}\rangle=\frac{1}{\sqrt{2}}(\sigma_{i}\otimes I_{3})(|0x'\rangle+|1y'\rangle),~~~j=1,2,3,\nonumber\\
&& |\psi_{4}\rangle=\frac{1}{\sqrt{2}}(|0\rangle+i|1\rangle)\otimes|z'\rangle,\nonumber\\
&& |\psi_{5}\rangle=\frac{1}{\sqrt{2}}(i|0\rangle+|1\rangle)\otimes|z'\rangle.
\end{eqnarray}
It can be directly verified that the two UMEBs (\ref{UMEB1'}) and (\ref{UMEB2'p}) satisfy the condition (\ref{mu23}).

As another example we choose
\begin{eqnarray}\label{exam2}
&&\theta_{1}=\frac{4\pi}{3},\,\theta_{2}=\pi,\, \theta_{3}=0,\theta_{4}=\pi,\theta_{5}=0,\theta_{6}=\pi,\nonumber\\
&&\theta'_{1}=\frac{\pi}{3},\theta'_{2}=\frac{\pi}{6}.
\end{eqnarray}
The corresponding unitary matrix $W$ and $S$ are of the form,
\begin{equation}
W=1/\sqrt{3}\begin{pmatrix}
\frac{-1-\sqrt{3}i}{2}&-i&-1\\
-1&\frac{\sqrt{3}-i}{2}&1\\
1&-i&-1\\
\end{pmatrix},
\end{equation}
\begin{equation}\label{condis'''}
S=\frac{1}{\sqrt{2}}\begin{pmatrix}
\frac{1+\sqrt{3}i}{2}&\frac{\sqrt{3}+i}{2}\\
\frac{-\sqrt{3}+i}{2}&\frac{1-\sqrt{3}i}{2}\\
\end{pmatrix}.
\end{equation}
The basis $\{|a\rangle, |b\rangle\}$ in $\mathbb{C}^{2}$ and the basis $\{|x'\rangle, |y'\rangle,
|z'\rangle\}$ in $\mathbb{C}^{3}$ are given by
\begin{eqnarray}\label{x,y,z,a,b}
&&|x'\rangle=\frac{1}{\sqrt{3}}(\frac{-1-\sqrt{3}i}{2}|0'\rangle-|1'\rangle+|2'\rangle)\nonumber,\\
&&|y'\rangle=\frac{1}{\sqrt{3}}(-i|0'\rangle+\frac{\sqrt{3}-i}{2}|1'\rangle-i|2'\rangle)\nonumber,\\
&&|z'\rangle=\frac{1}{\sqrt{3}}(-|0'\rangle+|1'\rangle-|2'\rangle),\nonumber
\end{eqnarray}
and
\begin{eqnarray}\label{x,y,z,a,b}
&&|a\rangle=\frac{1}{\sqrt{2}}(\frac{1+\sqrt{3}i}{2}|0\rangle+\frac{-\sqrt{3}+i}{2}|1\rangle),\nonumber\\
&&|b\rangle=\frac{1}{\sqrt{2}}(\frac{\sqrt{3}+i}{2}|0\rangle+\frac{1-\sqrt{3}i}{2}|1\rangle).
\end{eqnarray}
Therefore, another UMEB that is mutually unbiased to the UMEB given by (2) is of the form,
\begin{eqnarray}\label{UMEB2'pp}
&& |\psi_{j}\rangle=\frac{1}{\sqrt{2}}(\sigma_{i}\otimes I_{3})(|0x'\rangle+|1y'\rangle),~~~j=1,2,3,\nonumber\\
&&|\psi_{4}\rangle=\frac{1}{\sqrt{2}}(\frac{1+\sqrt{3}i}{2}|0\rangle+\frac{-\sqrt{3}+i}{2}|1\rangle)\otimes|z'\rangle,\nonumber\\
&&|\psi_{5}\rangle=\frac{1}{\sqrt{2}}(\frac{\sqrt{3}+i}{2}|0\rangle+\frac{1-\sqrt{3}i}{2}|1\rangle)\otimes|z'\rangle.
\end{eqnarray}

We have presented a general way in constructing UMEBs in
$\mathbb{C}^{2}\bigotimes\mathbb{C}^{3}$ such that they are mutually unbiased.
Explicit examples are given for constructing a pair of mutually unbiased
unextendible maximally entangled bases, including the one in \cite{Bin Chen}
as a special case. Our approach may shed light in constructing more UMEBs that are
pairwise mutually unbiased in $\mathbb{C}^{2}\bigotimes \mathbb{C}^{3}$ or higher dimensional bipartite systems.

\end{document}